\documentclass[aps,prd,nofootinbib, twocolumn, superscriptaddress]{revtex4-2}

\usepackage{amsmath,amssymb, amsthm,amstext}
\usepackage{natbib}
\usepackage{graphicx}
\usepackage{color}
\usepackage{array, enumerate}
\usepackage{bm}
\usepackage{multirow}
\usepackage[breaklinks,colorlinks,citecolor=blue]{hyperref}
\usepackage{braket}
\usepackage{txfonts}
\usepackage{physics}
\usepackage{xcolor}

\usepackage{soul}
\setulcolor{magenta}
\setul{-2.5pt}{1.5pt}

\def\be{\begin{equation}}
\def\ee{\end{equation}}
\def\bw{\begin{widetext}}
\def\ew{\end{widetext}}
\def\ba{\begin{aligned}}
\def\ea{\end{aligned}}
\def\bpm{\begin{pmatrix}}
\def\epm{\end{pmatrix}}

\begin{document}

\title{Gravitational waves from superradiant cloud level transition}
\author{Si-Tong Peng}
\email{pengsitong21@mails.ucas.ac.cn}
\affiliation{International Centre for Theoretical Physics Asia-Pacific, University of Chinese Academy of Sciences, 100190 Beijing, China}
\author{Jun Zhang}
\email{zhangjun@ucas.ac.cn}
\affiliation{International Centre for Theoretical Physics Asia-Pacific, University of Chinese Academy of Sciences, 100190 Beijing, China}
\affiliation{Taiji Laboratory for Gravitational Wave Universe (Beijing/Hangzhou), University of Chinese Academy of Sciences, 100049 Beijing, China}

\begin{abstract}

Ultralight boson clouds can form around black holes in binaries through superradiance, and undergo resonant level transitions at certain orbit frequencies. In this work, we investigate the gravitational waves emitted by the clouds during resonant level transitions, and forecast their detectability with future gravitational wave observations. We find that, for scalar fields of mass around $10^{-12}$ eV, clouds in stellar mass black hole binaries can radiate gravitational waves around $0.1$ Hz during hyperfine level transition, that could be detected with future gravitational wave detectors such as Big Bang Observer(BBO), but at a very low event rate. We also consider the clouds in intermediate mass black hole binaries, which can emit milli-Hz gravitational waves during hyperfine level transition. The resulting gravitational waves, however, can be hardly detected with Laser Interferometer Space Antenna(LISA)-like detectors.

\end{abstract}
\maketitle

\section{Introduction}

Ultralight bosons are predicted in many fundamental theories~\cite{Arvanitaki:2009fg}, and could constitute a portion or all of dark matter~\cite{Turner:1983he,Press:1989id,Hu:2000ke,Amendola:2005ad,Schive:2014dra,Hui:2016ltb}. Such ultralight bosons, if exist, could extract angular momentum from a rotating black hole through superradiance mechanism, forming a cloud around the black hole~\cite{Brito:2015oca, Arvanitaki:2014wva}. The eigenstates of the cloud are similar to those of a hydrogen atom, and hence the cloud and its host black hole is also called a gravitational atom. While the superradiance process will eventually stops as the black hole spins down, the cloud can exit for a relatively longer time, and radiate quasi-monochromatic gravitational waves (GWs), providing a smoking gun for the existence of the ultralight bosons~\cite{Brito:2017wnc, Brito:2017zvb, Tsukada:2018mbp, Isi:2018pzk, Palomba:2019vxe,Tsukada:2020lgt,Ng:2020ruv,LIGOScientific:2021rnv, Yuan:2022bem, Miller:2025yyx}.

Such superradiant clouds can also form around black holes in binary systems. In this case, the cloud can manifest rich dynamical phenomena~\cite{Baumann:2018vus,Baumann:2019ztm,Berti:2019wnn,Ikeda:2020xvt,Choudhary:2020pxy,Tong:2022bbl,Baumann:2021fkf,Guo:2024iye,Liu:2024mzw, Guo:2025ckp}, and play an important role in binary evolution~\cite{Cardoso:2011xi,Ferreira:2017pth,Zhang:2018kib,Zhang:2019eid,Baumann:2022pkl,Tomaselli:2023ysb,Cao:2023fyv, Tomaselli:2024bdd, Cao:2024wby,Takahashi:2024fyq, Zhu:2024bqs,Tomaselli:2024dbw,Boskovic:2024fga,Guo:2023lbv,Guo:2025ckp}. In particular, it is shown that a cloud that is initially in one of the eigenstates can resonantly transit to another state due to the gravitational perturbation of the companion object in the binary~\cite{Baumann:2018vus}. The process of transition between eigenstates is called cloud level transition. Cloud level transition typically accompanies with efficient momentum transfer between the cloud and the orbit of
the binary, and could lead to interesting orbital dynamics such as floating orbits and sinking orbits~\cite{Zhang:2018kib,Tomaselli:2024bdd}. In addition to level transitions, clouds in black hole binaries can undergo ionization~\cite{Baumann:2021fkf}, mass transfer~\cite{Guo:2023lbv,Guo:2024iye} and even form common envelope. All these processes could alter orbital decay, leaving fingerprints in GWs emitted by the binaries. While the GWs from clouds around isolate black holes could be confused with other quasi-monochromatic GW signals, the dynamical signatures of clouds in binaries can certainly help breaking the degeneracy.

In this work, we investigate the GW signals that are generated during cloud level transitions: As different eigenstates typically have different multipole moments, one may expect that the multipole moments of the cloud vary with time as it transits from one state to another during the level transition. As a result, the cloud will radiate GWs in additional to those emitted off the level transition. We study these GW signals in details, and forecast their detectability with future GW detectors. 

This paper is organized as follows: We start with reviewing the basics of cloud level transitions in Sec.~\ref{sec:transtion}. Then we calculate the level transition GW signals from a cloud consist of a complex field in Sec.~\ref{sec:GW}, and discuss their detectability in Sec.~\ref{sec:det}. Sec.~\ref{sec:con} devotes to conclusion and discussion.

\section{Cloud Level Transition}\label{sec:transtion}

In this section, we first review the basics of cloud level transitions that have been studied in Refs.~\cite{Baumann:2018vus,Baumann:2019ztm,Tomaselli:2024bdd}. We can start with a Kerr black hole of mass $M$ and dimensionless spin $a$, and a test complex scalar field $\Psi$ of mass $\mu$ in the black hole background.\footnote{The reason we consider a complex scalar field instead of a real scalar field is that, the cloud of the real scalar field constantly radiates GWs, and we would like to focus on GWs caused by cloud level transition. The obtained results, however, can be extended to the case of real scalar fields straightforwardly.} When $\alpha \equiv G M \mu < 1$, the scalar field may experience a superradiant instability, and form a cloud around the black hole. The instability manifests in the eigenfrequencies of the bound states of $\Psi$, which are shown to be a set of discrete complex number, and can be labelled by three quantum number $n$, $\ell$ and $m$. In particular, for $\alpha \ll 1$, the eigenfrequency can be estimated by
\be
\ba
{\rm Re} \omega_{n \ell m} &= \mu \left[ 1 - \frac{\alpha^2}{8 n^2} - \frac{\alpha^4}{8 n^4} + \frac{(2\ell-3n+1) \alpha^4}{n^4 (\ell+1/2)} \right. \\
& \quad \quad \left. + \frac{2 m a \alpha^5}{n^3 \ell (\ell+1/2) (\ell+1)}  + {\cal O} (\alpha^6) \right], \\
{\rm I m} \omega_{n \ell m} &= 2 \tilde{r}_+  C_{n\ell m}(\alpha)   \left(m\Omega_H - {\rm Re} \omega_{n\ell m}\right) \alpha^{4\ell+5},
\ea
\ee
where $\tilde{r}_+ = 1+\sqrt{1-a^2}$, $\Omega_H = a/2 GM \tilde{r}_+$ and 
\be
\ba
    C_{n \ell m} = & \frac{2^{4\ell+2}(2\ell+n+1)!}{(n+\ell+1)^{2\ell+4}n!} \left(\frac{\ell!}{(2\ell)!(2\ell+1)!}\right)^2 \\
  &  \times \prod^\ell_{j=1} \left( j^2(1-a^2)+(m a - 2 \tilde{r}_+ \alpha)^2 \right) \, .
     \ea
\ee
In the non-relativistic limit, it is convenient to consider the ansatz
\be
\Psi(t, \mathbf{r}) = \frac{1}{\sqrt{\mu}} \psi(t, \mathbf{r}) e^{-i\mu t} \, ,
\ee
where $\psi(t, \mathbf{r})$ is a complex scalar field that varies on a timescale much longer than $\mu^{-1}$. With this ansatz, the Klein-Gordon equation satisfied by $\Psi$ reduces to the Schr\"odinger equation,
\be
\label{eq:sch}
    i \frac{\partial}{\partial t} \psi(t, \mathbf{r}) = \left[ -\frac{1}{2 \mu} \nabla^2 - \frac{\alpha}{r} + {\cal O}(\alpha^2) \right] \psi(t, \mathbf{r}) \, .
\ee
To the leading order of $\alpha$, the eigenstates of Eq.~\eqref{eq:sch} are given by the eigen-wavefunctions of the hydrogen atom,
\be
\psi_{n \ell m} = e^{-i \epsilon_{n\ell m}  t} R_{n \ell}(r) Y_{\ell m}(\theta,\phi)\, ,
\ee
where $ \epsilon_{n \ell m} = \omega_{n \ell m} - \mu$ and $Y_{\ell m}$ is the spherical Harmonics. Given the energy stress tensor of the scalar field, the energy density of the cloud is dominated by $\mu  |\psi|^2$, and therefore $\psi$ can be normalized as $\int {\rm d}^3 x\, |\psi|^2 = M_{\rm c}/\mu$ with $M_{\rm c}$ being the mass of the cloud. For a given eigenstate $\psi_{n\ell m}$, the density profile is peaked around $n^2 r_b$ with $r_b = G M/\alpha^2$ being the Bohr radius. 

Now we consider that the black hole and its cloud is companied by another compact object of mass $M_*$, forming a binary system. The companion will perturb the cloud, introducing a tidal potential $V_*$ in the rest frame of the spinning black hole,
\be
\label{eq:schv}
    i \frac{\partial}{\partial t} \psi(t, \mathbf{r}) = \left[ -\frac{1}{2 \mu} \nabla^2 - \frac{\alpha}{r} + V_*(t,r) \right] \psi(t, \mathbf{r}) \, .
\ee
To the Newtonian order, the tidal potential induced by the companion can be written as
\be
\ba
V_*(t,r) = -GM_*\mu \sum_{\ell_* \geq 2} \sum_{\abs{m_*} \leq \ell_*} & \frac{4 \pi}{2l_*+1}  Y^*_{\ell_*m_*}\left(\theta_*,\phi_*\right)  \\
&\times Y_{\ell_*m_*}\left(\theta,\phi \right) F(r,\, r_*) ,
\ea
\ee
where $(r_*,\theta_*,\phi_*)$ are the coordinates of the companion, and
\be
F(r,\, r_*)=
\begin{cases}
\dfrac{r^{\ell_*}}{r_*^{\ell_*+1}}\Theta(r_*-r)+\dfrac{r_*^{\ell_*}}{r^{\ell_*+1}}\Theta(r-r_*)&\text{for }\ell_*\ne1\,,\\[12pt]
\biggl(\dfrac{r_*}{r^2}-\dfrac{r}{r_*^2}\biggr)\,\Theta(r-r_*)&\text{for }\ell_*=1\,,
\end{cases}
\ee
with $\Theta$ being the step function. To study the cloud evolution under the perturbation of the companion, we first consider the state in the interaction picture,
\be
\ket{\psi} \equiv \frac{\mu^{1/2}}{M_c^{1/2}} e^{i H_0 t} \psi(t) \,,
\ee
where $H_0$ is the diagonal and real part of the Hamiltonian. Generally, a cloud can be expressed as the superposition of the eigenstates
\be\label{eq:cloud}
\ket{\psi} = \sum_i \bar{c}_i(t)\, \ket{n_i \ell_i m_i} \, .
\ee
Substituting Eq.~\eqref{eq:cloud} into Eq.~\eqref{eq:schv} leads to
\be
\label{eq:ci}
    i \dot{\bar{c}}_i(t) = -i\, \Gamma_i \bar{c}_i + \sum_{i \neq j} \eta_{ij} \,e^{i \epsilon_{ij} t - i m_{ij} \phi_*(t)}\, \bar{c}_j(t) \,,
\ee
where the dot denotes the derivative with respect to $t$, $\Gamma_i \equiv - {\rm Im} \omega_{n \ell m}$ is the decay rate of $\ket{n_i \ell_i m_i}$, $\epsilon_{ij} \equiv \epsilon_i - \epsilon_j$ is the energy difference between the two states, and
\be
\eta_{ij} \equiv \mel{n_i \ell_i m_i}{V_*}{n_j \ell_j m_j} e^{ i\, m_{ij}\, \phi_*(t) } 
\ee
is the coupling strength with $m_{ij} \equiv m_i-m_j$.

Eq.~\eqref{eq:ci} indicates that a cloud that is initially in one of the eigenstates can transit to another state resonantly under the perturbation of the companion. For circular orbits with no inclination, the resonant transition typically involves only two states at a time. To be concrete, we can consider the transition from $\ket{n_1\ell_1m_1}$ to $\ket{n_2\ell_2m_2}$, assuming $\ket{n_1\ell_1m_1}$ is saturated such that $\Gamma_1 = 0$ while $\ket{n_2\ell_2m_2}$ is an unstable state with $\Gamma_2 > 0$. Then wavefunction of cloud is given by
\be\label{cloud2states}
\ket{\psi} = \bar{c}_1(t) \ket{n_1\ell_1m_1} + \bar{c}_2(t) \ket{n_2\ell_2m_2} \,.
\ee
The resonant transition happens when the orbital frequency approaches to the resonant frequency $\Omega_0 = \epsilon_{12}/ m_{12} $. As approaching to the resonant orbit, the orbital phase can be expanded as
\be
\phi_*(t) \simeq \Omega_0 t + \frac12 \gamma_0 t^2 \, .
\ee
For example, assuming the orbital decay is dominated by GW radiation, we have
\be
\gamma_0^{\rm GW} = \frac{96}{5} \frac{q}{(1+q)^{1/3}} (G M\Omega_0)^{5/3} \Omega_0^2 \,,
\ee
where $q= M_*/M$ is the mass ratio.
To solve $\bar{c}_1$ and $\bar{c}_2$, we further introduce 
\be
\ba
&\bar{c}_{1} = e^{\frac{i}{2}\epsilon_{12}t - \frac{i}{2} m_{12} \phi_*} c_{1} \,, \\
&\bar{c}_{2} = e^{ \frac{i}{2} m_{12} \phi_* -\frac{i}{2}\epsilon_{12}t }  c_{2} \,,
\ea
\ee
Eq.~\eqref{eq:ci} can be then written as
\be
\ba\label{eq:ci2}
    &i\dot{c}_1 =  \frac{m_{12} \gamma_0 }{2} t c_1 + \eta_{12} c_2 \,, \\
    &i\dot{c}_2 =  \eta_{12} c_1 - \left(\frac{m_{12} \gamma_0 }{2} t + i \Gamma_2\right) c_2 \,. \\
\ea
\ee
With $c_1 = 1$ and $c_2 = 0$ as $t$ approaches $-\infty$, the solution of Eq.~\eqref{eq:ci2} is known as,
\be
\ba\label{c1c2sol}
    &c_1(\tau) = e^{i \frac{\pi}{4} - \frac{\Gamma_2 t}{2} - \frac{\pi Z}{4}} {\cal D}_{iZ}\left[e^{i \frac{3\pi}{4}}\left(\sqrt{\abs{m_{12}} \gamma_0} \,t - \frac{i\Gamma_2}{\sqrt{\abs{ m_{12}} \gamma_0}}  \right)\right], \\
    &c_2(\tau) = e^{-\frac{\Gamma_2 t}{2} - \frac{\pi Z}{4}} \sqrt{Z} {\cal D}_{iZ-1}\left[e^{i \frac{3\pi}{4}}\left( \sqrt{\abs{m_{12}} \gamma_0} \,t- \frac{i\Gamma_2}{\sqrt{\abs{ m_{12}} \gamma_0}} \right)\right], \\
\ea
\ee
where ${\cal D}_{a} (x)$ is the parabolic cylinder function with $Z \equiv \eta_{12}^2/ \left| m_{12}\right|\gamma_0$.

Depending on the involved states, there are three types of resonances: Bohr resonance ($n_1 \neq n_2$), fine resonance ($n_1 = n_2$ and $\ell_1 \neq \ell_2$), and hyperfine resonance ($n_1=n_2$, $\ell_1= \ell_2$, and $m_1\neq m_2$).
Taking the hyperfine level transition from $\ket{211}$ to $\ket{21-1}$ for example, when $\ket{211}$ saturates, we have $a \simeq 4\alpha/\left(1+4\alpha^2\right)$, $\Omega_0 \simeq \alpha^7/3GM$, $\eta \simeq q \alpha^{11}/GM$, $\Gamma_2 \simeq 5 \alpha^{10}/96GM $, and $\gamma_0^{1/2} \propto q^{1/2} \alpha^{77/6}/G M$.
For comparable mass ratio binaries with $\alpha \ll 1$, we expect that $\Omega_0 \gg \Gamma_2 \gg \eta \gg \gamma_0^{1/2}$. Note that while Eq.~\eqref{eq:schv} and the resulting level transitions precisely describe the cloud's response to the companion's tidal perturbation, the expansion of cloud into its bound states only works when the orbital separation is much larger than $r_b$. As the orbital separation becomes comparable to $r_b$, other states, such as unbound states and states bounded by the two black holes could be excited. In particular, tidal disruption may happen when the orbital separation is around $2.6 r_b$ for comparable mass ratio binaries~\cite{Baumann:2018vus}. For the hyperfine level transition discussed in this work, the orbital separation is larger than $38 r_b$, therefore, the tidal disruption is not relevant, and the cloud deformation is well described by Eq.~\eqref{eq:schv} and Eq.~\eqref{eq:cloud}.

\section{GW Radiation from Cloud Level Transition}\label{sec:GW}

In this section, we investigate the GW radiation during the cloud resonant level transition. During the transition, the cloud oscillates with a frequency of $\sim |\Delta \epsilon|$. Therefore, the wavelength of the resulting GWs is of $|\Delta \epsilon|^{-1} \sim \mu^{-1}\alpha^{-p}$ with $p=2$, $4$ and $5$ for Bohr, fine, and hyperfine resonances respectively. Different from cloud annihilation, the wavelength of GWs from cloud level transitions is thus much larger than the size of the cloud, i.e., $\Delta \epsilon \, r_b \sim  \alpha^{p-1} \ll 1$ for $\alpha \ll 1$. In this case, the GW radiation can be calculated with the multipole expansion.

In the Newtonian limit, the energy-stress tensor of the cloud is dominated by the energy density $\rho \approx \mu |\psi|^2$, and hence the multipole of cloud can be estimated as
\be
    I_{lm} = \frac{16 \pi \mu}{(2l+1)!!} \left[\frac{(l+1)(l+2)}{2l(l-1)}\right]^{1/2} \int {\rm d}^3x\,  r^l  |\psi|^2 Y^*_{lm} \,.
\ee
Then the corresponding GW radiation in the Cartesian coordinates can be calculated in the transverse-traceless gauge,
\be\label{hij}
 h_{ab}  = \frac{G M_c}{r} \sum_{l=2}^{\infty} \sum_{m=-l}^l u_{lm}\left(t' \right) \, \mathbf{T}^{lm}_{ab}(\theta, \phi),
\ee
where the subscription $a$, $b$ denote the Cartesian bases, $u_{lm} = \tfrac{{\rm d}^l}{{\rm d}t^l} I_{lm}$, $\mathbf{T}^{lm}_{ab}(\theta, \phi) = \left(2\frac{(l-2)!}{(l+2)!}\right)^{1/2} r^2 \Lambda_{ab, a'b'}(\hat{n}) \partial_{a'} \partial_{b'} Y_{lm} (\theta, \phi)$, with $\Lambda_{ab, a'b'}$ being the projection tensor, $r$ being the distance between the observer and the cloud, $\hat{n}$ being the direction of the propagation of GWs and $t' = r-t$. With ansatz \eqref{eq:cloud}, the multipole can be written as
\be
I_{lm} = \sum_{i, j} c_i^*(t) c_j(t) e^{i m_{ij} \phi_*(t)} r_b^l C^{lm}_{ij}  \,,
\ee
where
\be\label{eq:Clmij}
C^{lm}_{ij} \equiv \frac{16 \pi }{(2l+1)!!} \left[\frac{(l+1)(l+2)}{2l(l-1)}\right]^{1/2}  \mel{n_i \ell_i m_i}{r^l r_b^{-l} Y^*_{lm} }{n_j \ell_j m_j}  \,, 
\ee
is time independent. The integral in $C^{lm}_{ij}$ indicates that $C^{lm}_{ij}$ is non-trivial only if $m_i-m_j+m=0$, $\abs{\ell_i-\ell_j} \le l \le \ell_i+\ell_j$ and $\ell_i+\ell_j+l= 2p$ for $p \in \mathbb{Z}$. 

It is claimed in Ref.~\cite{Brito:2017zvb} that for GWs from cloud annihilation, the radiation power obtained with the flat space approximation may differs from that obtained with the numerical computations by one or two orders of magnitude for large $\alpha$. In that case, the GW wavelength is of $\mu^{-1} = GM \alpha^{-1}$ and is comparable to the size of the black hole for $\alpha > 0.2$, therefore, the effects from the black hole background should be taken into account. In our case, the GW wavelength is much larger than the size of the black hole. Specifically, for GWs from the hyperfine level transition that we focus on in the next section, the GW wavelength is of $\alpha^{-7} GM$, and the corresponding small parameter in the post-Newtonian expansion is $(GM\Omega_0)^{2/3} \simeq 0.006$ for $\alpha = 0.42$. Therefore, we expect the Newtonian multipole expansion works well in our case.

Now we focus on the two-state model. When calculating $u_{lm}$, the time derivative on $c_i$ can be replaced utilizing Eq.~\eqref{eq:ci2}. By doing so, we can find that terms in $u_{lm}$ can be arrange in the powers of $\eta_{12}$, $\gamma_0 t$, $\Gamma_2$ and $\Omega_0$ (which arises from acting time derivative on $e^{i m_{ij} \phi_*}$ in $I_{lm}$). Given the hierarchy of these quantities, we can conclude that $u_{l m}$ is dominated by terms proportional to $\Omega_0^{l}$. As a result, the cloud level transition GW are dominated by
\be
 h_{ab} \simeq \frac{2GM_c}{r} (\Omega_0 r_b)^l \, {\rm Re }\left[(i m_{12})^l c_1^* c_2 e^{i m_{12} \phi_*}  C^{lm}_{12}\right]  \, \mathbf{T}^{lm}_{ab}(\theta, \phi),
\ee
where $l$ corresponds to the lowest multipole that $l \le 2$ and $C^{lm}_{ij} \neq 0$, and $c_1$, $c_2$ and $\phi_*$ are functions of $t'$.

\begin{figure}
    \centering
     \includegraphics[width=0.97\linewidth]{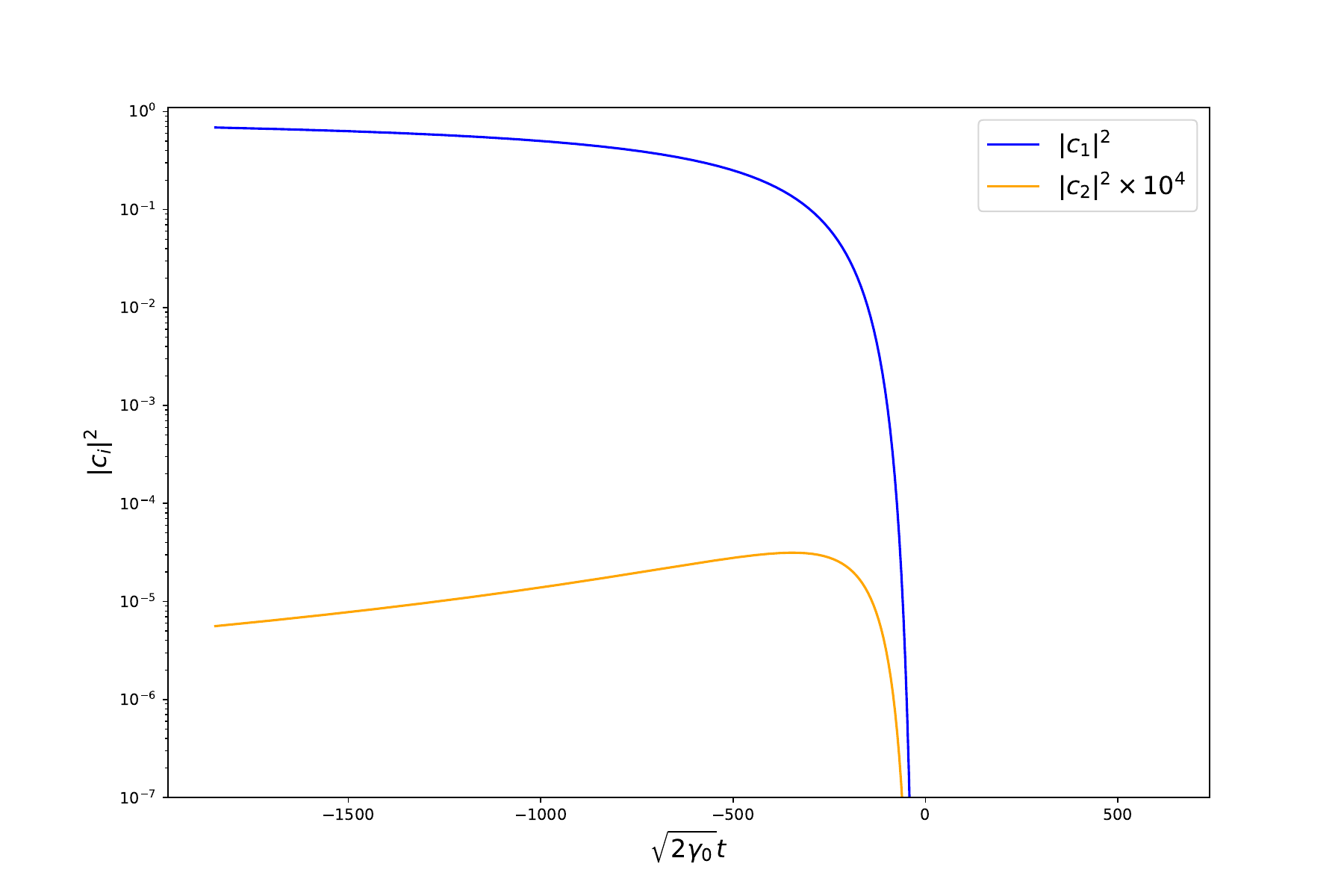} 
    \includegraphics[width=0.97\linewidth]{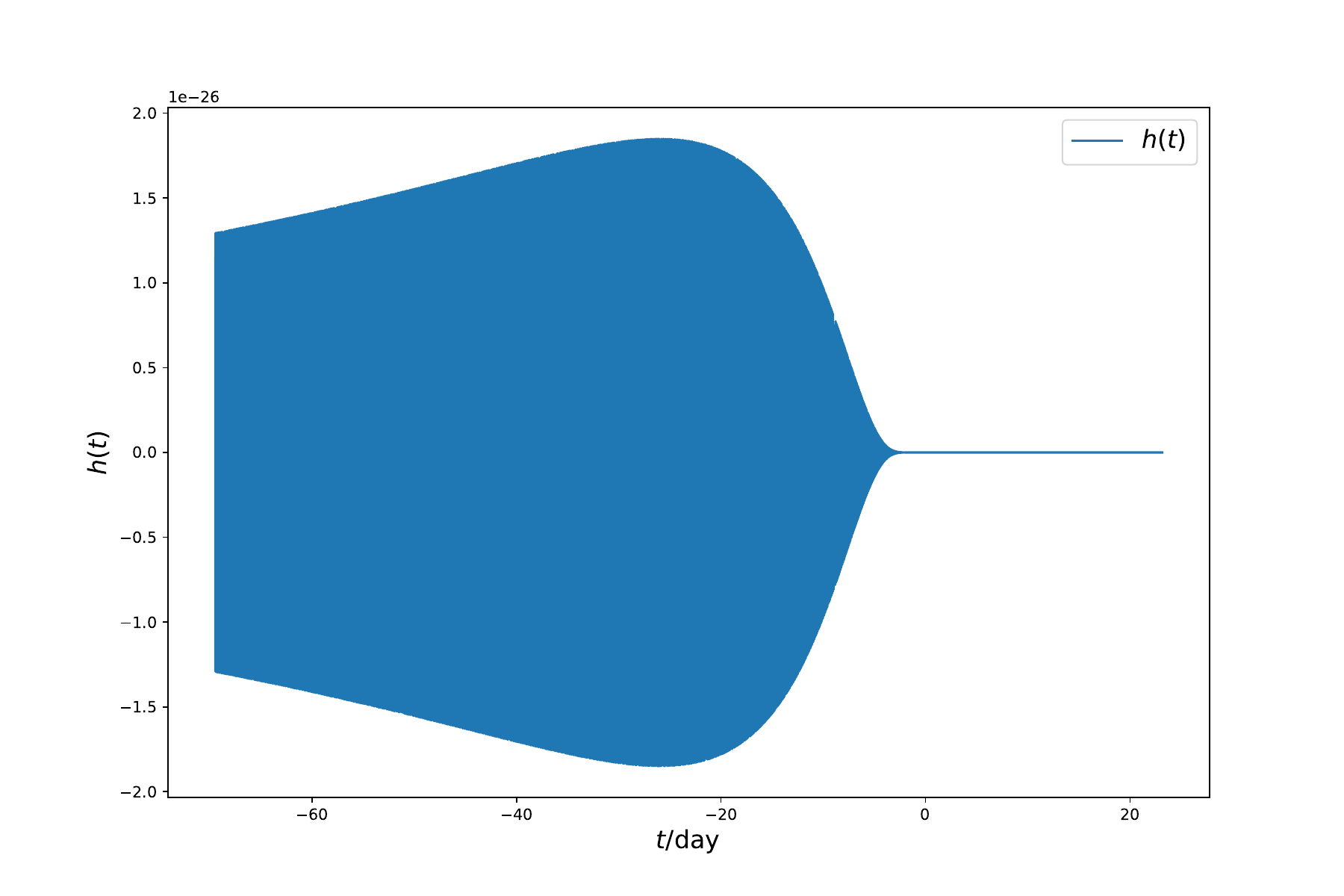}  \\
     \caption{GWs from cloud level transition between $\ket{21\pm1}$. We consider a scalar field with $\mu = 10^{-12} {\rm eV}$, and a black hole binary at redshift $z=0.001$. The masses of the black holes are $40 M_\odot$ and $20 M_\odot$, i.e., $q=0.5$. We assume the primary black hole is dressed with a cloud of mass $M_c = 4 M_\odot$ that was initially in $\ket{211}$, and is transferring to $\ket{21-1}$ as the binary approaching to the resonant orbit at $t=0$. The upper panel shows the evolution of the cloud, while the lower panel shows the resulting GW signals in the time domain.}
    \label{fig:signal}
\end{figure}

Assuming the cloud is initially in $\ket{211}$, it may experience hyperfine level transition, i.e., from $\ket{211}$ to $\ket{21-1}$. Then the level transition GW are dominated by quadrupole radiation with $l=2, m=\pm 2$. The GW strain, averaged over all possible inclination angle, is
\be
\ba
h^{\rm H}_{211}  \approx & \frac{32 \sqrt{10 \pi} }{75} \frac{G M_{\rm c}}{r} \alpha^8 a^2  {\rm Re} \left[ c_1^*(t') c_2(t') e^{-2i\phi_*(t')} \right]   \,.
\ea
\ee
In Fig.~\ref{fig:signal}, we show a typical GW signal from the $\ket{211}$ to $\ket{21-1}$ hyperfine level transition in the time domain, where we assume $M=40 M_\odot$, $\mu= 10^{-12} {\rm eV}$ and $q=0.5$. In this case, the signal lasts nearly a hundred days, and reaches its maximum around 20 days before the binary reaches the resonant orbit, namely $t=0$ in the plot.

\begin{figure}
    \centering
    \includegraphics[width=0.95\linewidth]{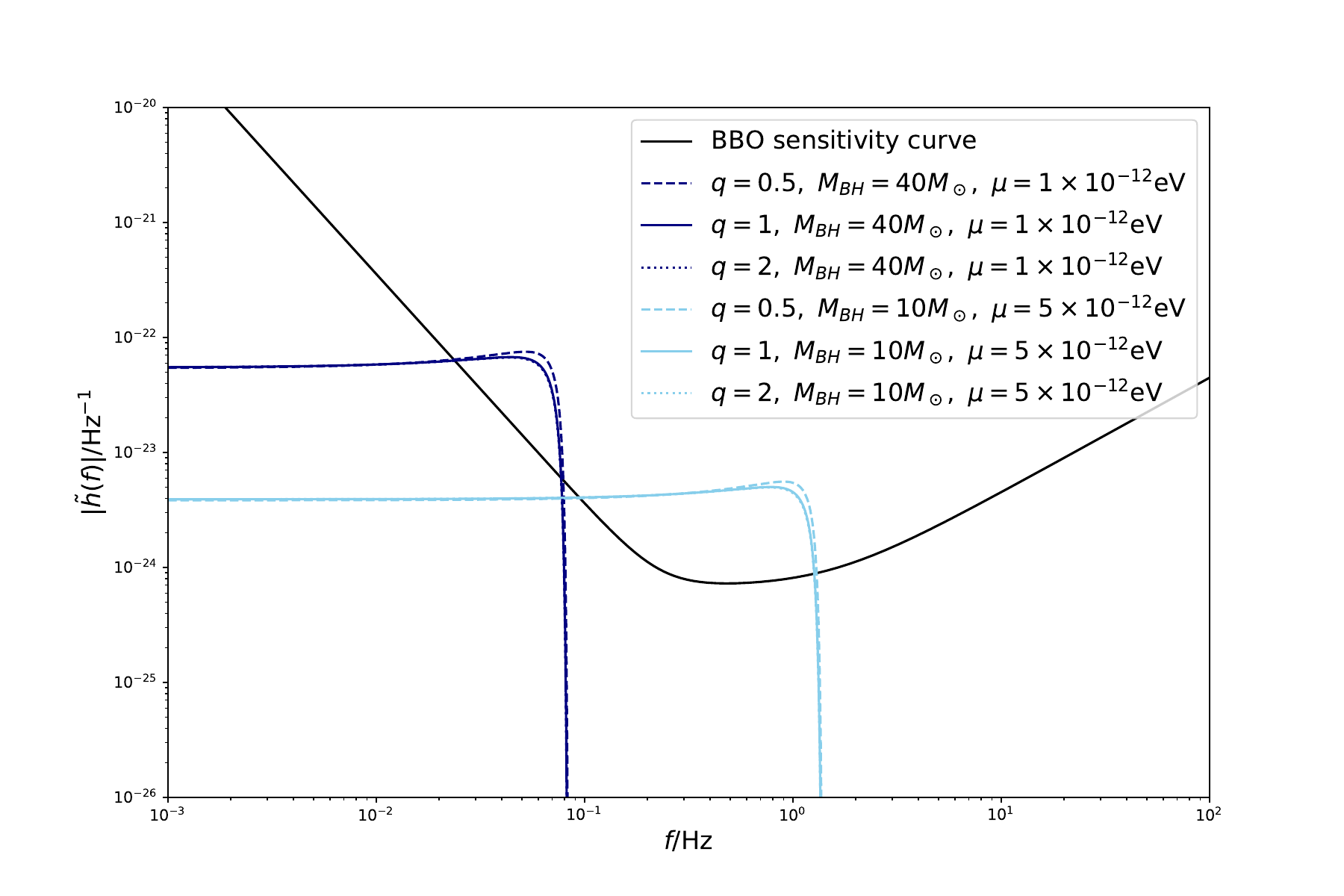}  \\
     \caption{Frequency domain signals of GWs from cloud level transition. We considered the $\ket{211}$ to $\ket{21-1}$ hyperfine level transition happen in a black hole binary at redshift $z=0.001$. The dark (light) blue lines show the strain with $M=40 M_\odot$ and $\mu = 10^{-12} {\rm eV}$ ($M=10 M_\odot$ and $\mu = 5\times 10^{-12} {\rm eV}$). The dashed, solid, dotted lines denote $q = 0.5$, $1$, and $2$, respectively. We find that, for comparable mass ratio binaries, the mass ratio does not affect the strain significantly. The black solid line shows the sensitivity curve of BBO.}
    \label{fig:strain}
\end{figure}

Given the fact that $c_1^* c_2$ changes much lower than $\Omega_0^{-1}$, the characteristic frequency of GWs is determined by
\be\label{eq:f0}
\ba
f_0^{\rm H} = \frac{\Omega_0}{\pi} &\approx \frac{ \alpha^7 }{3\pi GM} \\
& \approx 138  \left( \frac{\alpha}{0.02}\right)^7 \left(\frac{M}{2\times 10^{-10} M_\odot}\right)^{-1} {\rm Hz} \\
 & \approx 0.1  \left( \frac{\alpha}{0.3}\right)^7 \left(\frac{M}{40 M_\odot}\right)^{-1} {\rm Hz} \\
 &\approx 5  \left( \frac{\alpha}{0.3}\right)^7 \left(\frac{M}{10^3 M_\odot}\right)^{-1} {\rm mHz} \,,
\ea
\ee
where we have assumed the black hole spin $a \approx 4\alpha$ for $\ket{211}$ being saturated. For stellar mass black holes, the hyperfine level transition GWs are in the deci-Hz band, while for intermediate mass black holes, the hyperfine level transition GWs are in the milli-Hz band. Moreover, if there are binaries of primordial black holes, the hyperfine level transition GWs can be in the kilo-Hz band. The GW strains in the frequency domain can be estimated using the stationary phase approximation,
\be
\ba\label{eq:hf}
 \left| \tilde{h}^{\rm H}_{211} (f) \right| \approx   \frac{16}{5} \left(\frac{2}{3}\right)^{2/3}  \frac{G M_{\rm c}}{r}  \frac{(1+q)^{1/6}}{q^{1/2}}  \alpha^{3}  a^{\frac{7}{6}}\frac{  \left| c_1^*(t_*) c_2(t_*) \right| }{f_0^{\rm H}}   \,,
\ea
\ee
where $t_* =  \pi ( f -  f_0^{\rm H}) /  \gamma_0$. In Fig.~\ref{fig:strain}, we show the strain of GWs from the $\ket{211}$ to $\ket{21-1}$ hyperfine level transition in the frequency domain, assuming black hole binaries at redshift $z = 0.001$. We find that, for stellar mass black hole binaries with comparable mass ratio, the mass ratio does not affect the strain significantly.
 
The fine level transition, i.e., from $\ket{211}$ to $\ket{200}$, can in principle result in GW radiations. Nevertheless, given the selection rules raise in $C_{ij}^{lm}$, the GW radiations are at least octupole, and hence are suppressed. For co-rotating systems, i.e., the black hole spin aligns with the orbital angular momentum, the Bohr level transition, i.e., from $\ket{211}$ to $\ket{100}$, does not result in significant GW radiations either. For contour-rotating system, Bohr level transition could happen between $\ket{211}$ and  $\ket{n \ell m}$ with $n>2$, resulting in quadrupole GW radiation.

One may have noted that the GWs discussed above are calculated in the rest frame of the host black hole, and should be modulated by its orbital motion. Assuming the binary is at rest with respect to the GW detector, the modulation in phase can be estimated by considering a monochromatic wave emitted by the cloud, and the detector sees a signal proportional to
\be
\ba
&\cos\left[ |\Delta \epsilon| \left( t + R \sin\Omega t \right) \right] \\
 = &\sum_{k=-\infty}^{\infty} J_k\left(|\Delta \epsilon| R  \right)^k  \cos\left[\left(|\Delta \epsilon| + k \Omega \right) t \right] \,,
\ea
\ee
where $J_k$ is the Bessel function and $R$ is the orbital radius of the black hole binary. During the level transition, we have $|\Delta \epsilon| R \sim \alpha^{(p+1)/3} \ll 1$, and hence most of the power is still in the state with $k=0$, indicating that the orbital motion modulation is negligible. Similarly, modulations such as Doppler shift and gravitational redshift are also negligible, because of the slow motion of the orbits and the weak gravitation of the companion.

For relatively large $\alpha$, cloud in $\ket{211}$ might be depleted, and the $\ket{322}$ state could sufficiently grow during astrophysical timescale. In this case, the cloud might also radiate GWs, that are dominated with the multipole of $l=2$, $m = \pm2$ ($\pm1$) during the $\ket{322}$ to $\ket{320}$ ($\ket{321}$) hyperfine transition, of $l=3$, $m = \pm1$ during the $\ket{322}$ to $\ket{311}$ fine transition, and of $l=3$, $m = \pm (2-m_2)$ during the $\ket{322}$ to $\ket{21m_2}$ Bohr transition.

\section{Detectability}
\label{sec:det}

In this section, we assess the detectability of GW signals from cloud level transitions with future GW detectors. There are several timescales that are relevant for the discussion: the superradiant growth timescale $\tau_{\rm sr}$, the orbit lifetime $\tau_{\rm d}$ after black hole formation, the orbit lifetime after level transition $\tau_{\rm lt}$, and the cloud lifetime $\tau_{\rm c}$. We shall consider comparable mass binaries, and focus on the hyperfine level transition from $\ket{211}$ to $\ket{21-1}$. In this case, the time scale of floating orbit is much shorter than the lifetime of the resonant orbit without cloud backreaction. Therefore, we shall assume the orbit decays by simply radiating GWs, and define $\tau_{\rm d}$ and $\tau_{\rm lt}$ accordingly. Comparing to clouds consist of a real field, a complex cloud around an isolated black hole could live a longer life, as it does not radiate monochromatic GWs. Nevertheless, we assume that the cloud has a lifetime same as a real cloud, so that the discussion can be extended to the case of real clouds. Under these assumptions, occurrence of a level transition typically requires $\tau_{\rm sr} \ll \tau_{\rm lt}$\footnote{Strictly speaking, $\tau_{\rm sr} \ll \tau_{\rm lt}$ is not necessary for the occurrence of level transition, providing $\tau_{\rm sr} \ll \tau_{\rm d}$ and $\tau_{\rm lt} < \tau_{\rm d}$. It is necessary only when $\tau_{\rm d} \sim \tau_{\rm lt}$. Nevertheless, for the transition from the $\ket{211}$-state to the $\ket{21-1}$-state,  $\tau_{\rm sr} \ll \tau_{\rm lt}$ indicates $\alpha < 0.42$ which roughly the condition for the superradiant growth.} and $\tau_{\rm lt} \ll \tau_{\rm c}$ such that a cloud can fully grow and does not deplete before the level transition. These requirements indicate $0.13 < \alpha < 0.42$. We expect the transition happens within a timescale
\be
\ba
\tau_{\rm lt} &= \frac{5}{256}\frac{(1+q)^{1/3}}{q} \frac{G M}{ \left(G M \Omega_0\right)^{8/3} }  \\
 &\approx  15\times \left(\frac{M}{30 M_\odot}\right)^{-5/3} \left(\frac{f}{0.1 {\rm Hz}}\right)^{-8/3} {\rm day}\\
  &\approx 10 \times \left(\frac{M}{30 M_\odot}\right) \left(\frac{\alpha}{0.3}\right)^{-16} \left(\frac{a}{0.88}\right)^{-8/3} {\rm day} \, ,
\ea
\ee
where we take the binary mass ratio $q=1$ in the last two lines. One should note that for $\alpha > 0.2$ relativistic corrections to level structure can become important for level structure.

\subsection{Single events}

GWs from cloud level transition are always accompanied with the GWs from the binary, which is also of the frequency $\Omega_0/\pi$,
\be
h_{\rm s} = h_{\rm bhb} + h_{\rm c} \,,
\ee
where $h_{\rm s}$, $h_{\rm bhb}$ and $h_{\rm c} $ denote the GW signals detected by the detector, emitted by the black hole binary, and emitted by the cloud during level transition respectively. The detectability of the level transition signals can be assessed by measurement accuracy of the cloud mass. Using the Fisher matrix, the error in measuring the cloud mass is
\be
\Delta M_{\rm c} \simeq |\left(\partial_{M_{\rm c}} h_{\rm s} | \partial_{M_{\rm c}} h_{\rm s} \right) |^{-1/2},
\ee
where the inner product is defined as
\be
\left( h_1 |  h_2 \right) = 4 {\cal R} \int {\rm d}f \frac{\Tilde{h}_1(f) \Tilde{h}_2^*(f)}{{\rm S}_{\rm n}(f)} \,,
\ee
with $\Tilde{h}$ denoting the Fourier transformation of $h$ and ${\rm S}_{\rm n}(f)$ denoting the noise spectrum of the GW detector. Recall that $h_{\rm c} \propto M_{\rm c}$, we have
\be
\frac{\Delta M_{\rm c}}{M_{\rm c}} \simeq \frac{1}{\rho_{\rm SNR}} \, ,
\ee
where $\rho_{\rm SNR} \equiv \left( h_{\rm c} |  h_{\rm c} \right)^{1/2}$ is the signal to noise ratio (SNR) of the cloud level transition GWs.

\begin{figure}
    \centering
    \includegraphics[width=\linewidth]{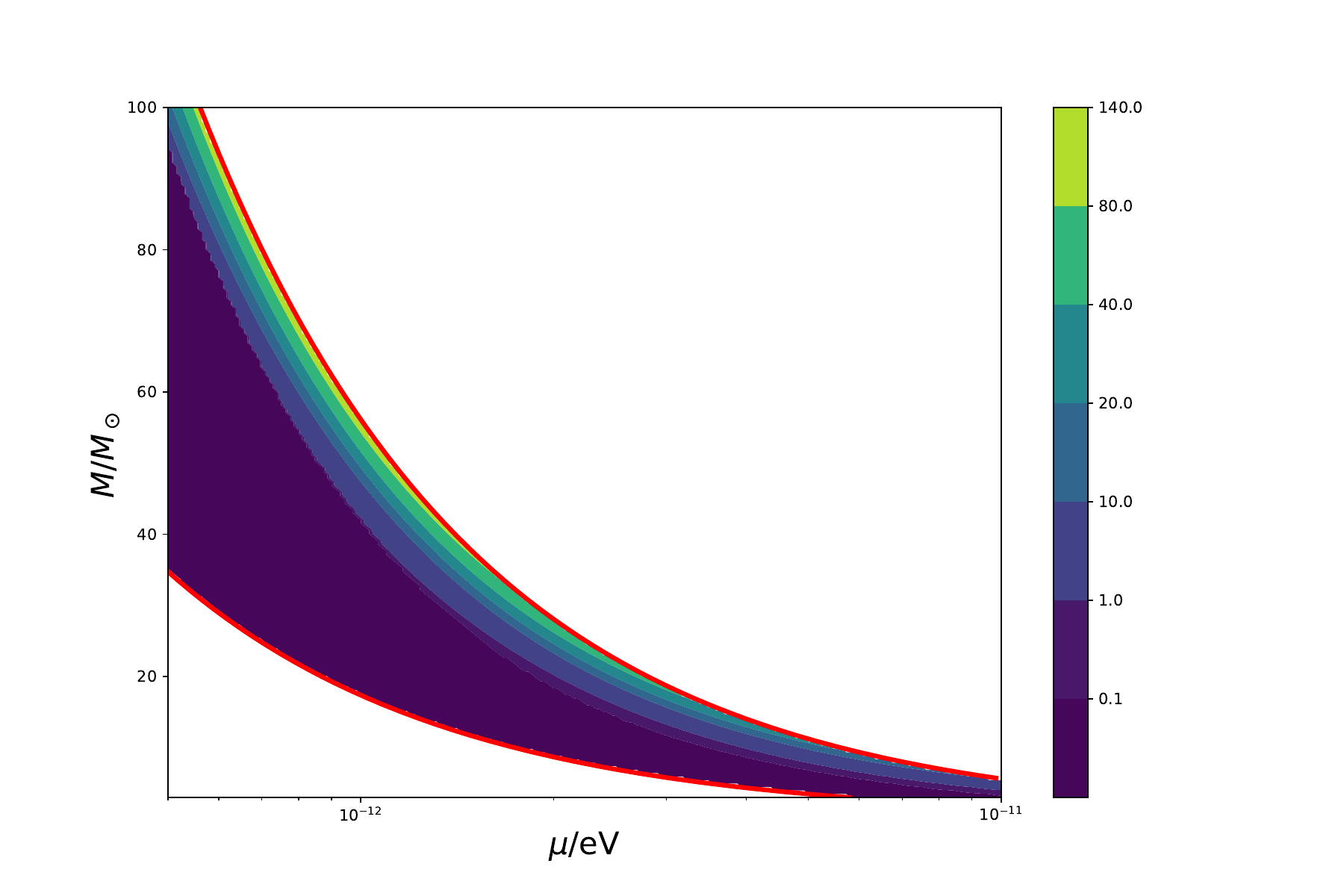} 
    \caption{SNR of cloud level transition GWs observed in the deci-Hz band. We consider the hyperfine cloud level transition from $\ket{211}$ to $\ket{21-1}$, assuming a scalar field of mass $\mu$ and equal mass binaries with black holes of mass $M$ at redshift $z=0.001$ and detected by a BBO-like GW detector. The cloud mass $M_{\rm c}$ is assumed to be $0.1M$. We are only interested in the region with $0.13<\alpha<0.42$ (bounded by the red lines) so that the clouds can fully grow and does not deplete before the level transition.}
    \label{fig:SNR_BBO}
\end{figure}

\begin{figure}
     \centering
     \includegraphics[width=\linewidth]{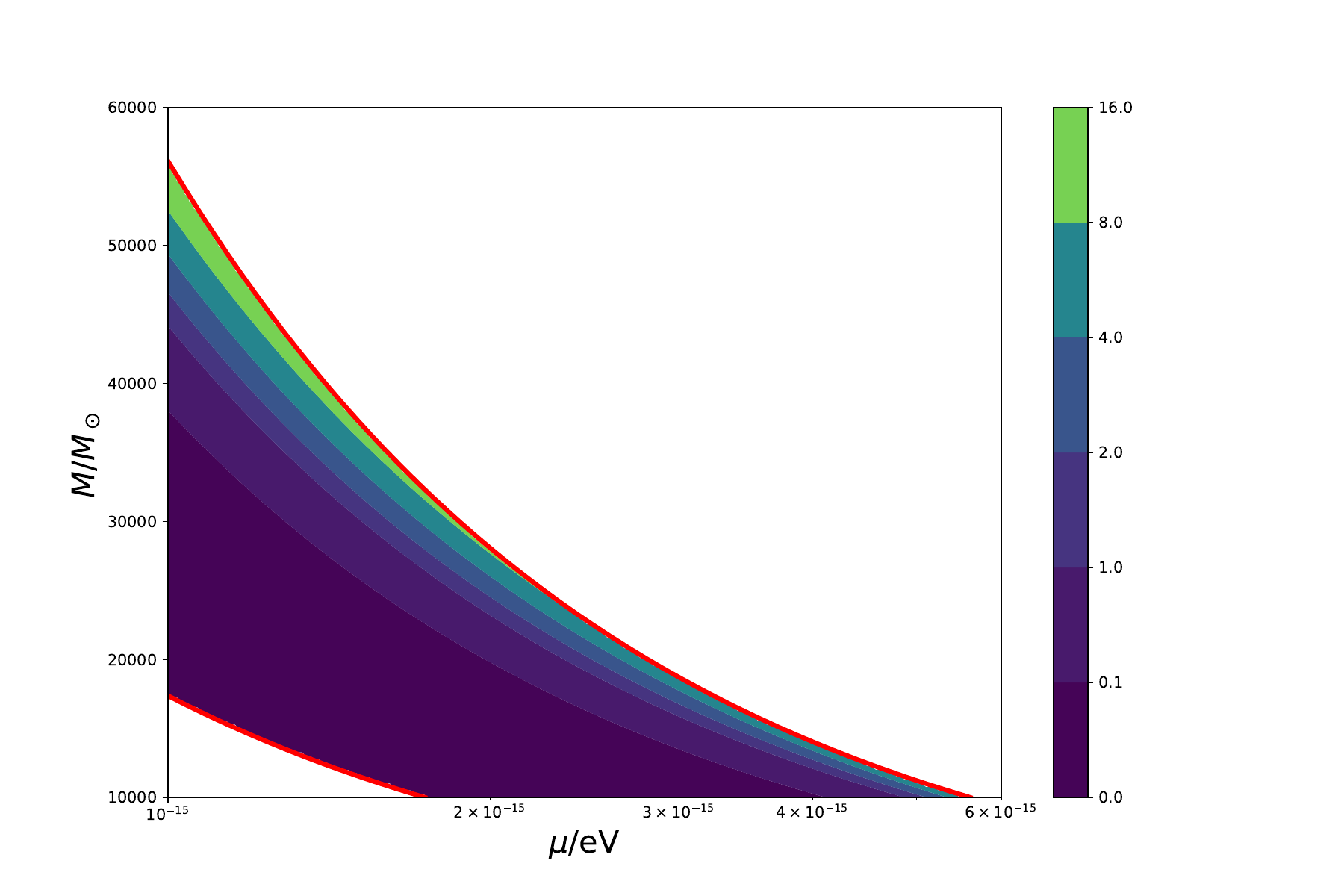}
          \caption{SNR of cloud level transition GWs observed in the milli-Hz band. We consider the hyperfine cloud level transition from $\ket{211}$ to $\ket{21-1}$, assuming a scalar field of mass $\mu$ and binaries with black holes of mass $M$ and $q=0.003$ at redshift $z=10^{-4}$ and detected by a LISA-like GW detector. The cloud mass $M_{\rm c}$ is assumed to be $0.1M$. The red lines show the region of $0.13<\alpha<0.42$ so that the clouds can fully grow and does not deplete before the level transition.}
     \label{fig:snrm}
 \end{figure}

We first consider stellar mass black hole binaries, the corresponding level transition GWs could be in the deci-Hz band. In Fig.~\ref{fig:SNR_BBO}, we show the SNR of the level transition GWs detected by BBO~\cite{Yagi:2011wg}, taking equal mass binaries at redshift $z = 0.001$ and clouds of mass $M_{\rm c} = 0.1 M$ as benchmarks. We are only interested in the region with $0.13 < \alpha < 0.42$ so that the clouds can fully grow and do not deplete before the level transition. We find that the GW signals typically have larger SNR for larger $\alpha$.  In addition, we consider intermediate mass black hole binaries, the corresponding level transition GWs are in milli-Hz band, cf. Eq.~\eqref{eq:f0}. Assuming a LISA-like detector~\cite{Robson:2018ifk}, the SNR of the level transition GWs is shown in Fig.~\ref{fig:snrm}.

Given the SNR of cloud level transition GWs, we shall further estimate the rate of detectable events. For a certain scalar field of mass $\mu$, the event rate can be estimated by integrating all binary black hole merger events with $0.13<\alpha<0.42$ and with $\rho_{\rm SNR} \ge 8$,
 \be
 R_{\rm lt} (\mu) =\int {\rm d} z  \int_{M_{\rm min}(z)}^{M_{\rm max}(z)} {\rm d} M  R_{\rm mrg} \left(M, t(z)\right) \frac{{\rm d} V_c}{{\rm d} z} \,,
 \ee
where $\frac{{\rm d} V_c}{{\rm d} z}$ is differential comoving volume, $R_{\rm mrg}$ is the merger rate, and we have assumed all binaries are equal-mass for simplicity, as the mass ratio does not affect the SNR for comparable mass ratio binaries (cf.~Fig.~\ref{fig:SNR_BBO}). The merger rate depends on the black hole formation history, and is computed following Ref.~\cite{Dvorkin:2016wac},
 \be
 \label{eq: merger rate}
 R_{{\rm mrg}}(M, t) = N \int_{\tau_{min}}^{\tau_{max}} d\tau_{\rm d} R_{\text{birth}}(M, t-\tau_{\rm d}) P_{\rm d}(\tau_{\rm d}) \, ,
 \ee
 where $R_{\rm birth} (M, t)$ is the black hole birth rate at time $t$ and is normalized with the observed rate of $10^{-7} \rm{Mpc}^{-3} \rm{yr}^{-1}$~\cite{LIGOScientific:2016kwr} to account for black holes that did not merge at all, $P_{\rm d}(\tau_{\rm d}) \propto \tau_{\rm d}^{-1}$ is the distribution of time delay distribution, $\tau_{max}$ is the cosmic time, and $\tau_{min}$ is $50\text{Myr}$ or $\tau_{\rm lt}$, whichever is larger. In principle, $R_{{\rm mrg}}(M, t)$ might depend on the mass of field $\mu$. However, in the case that considered above, it always has $\tau_{\rm lt} < 50\text{Myr}$, and thus $R_{{\rm mrg}}(M, t)$ does not depend on $\mu$. For concreteness, we consider black hole binaries formed via massive binary stars, in which case the black hole birth rate can be given by
 \be
 \ba
 &R_{\text{birth}}(M, t(z)) \\
 = &\int {\rm d} m_\star  \,  \text{SFR}(z) \, \text{IMF}(m_\star) \,  \delta(m_\star - g^{-1}(M,\mathcal{Z})).
 \ea
 \ee
 Here, ${\rm SFR} (z)$ is the star formation rate~\cite{Vangioni:2014axa}, ${\rm IMF}(m_\star) \propto m_\star^{-2.35}$ is the initial mass function~\cite{Salpeter:1955it}, and $g(m_\star,\mathcal{Z})$ is the relation between the remnant mass $M$ and the progenitor mass $m_\star$~\cite{Woosley:1995ip} and also depends on the redshift through the metallicity $\mathcal{Z}$~\cite{Belczynski:2016obo}. 

 \begin{figure}
     \centering
     \includegraphics[width=\linewidth]{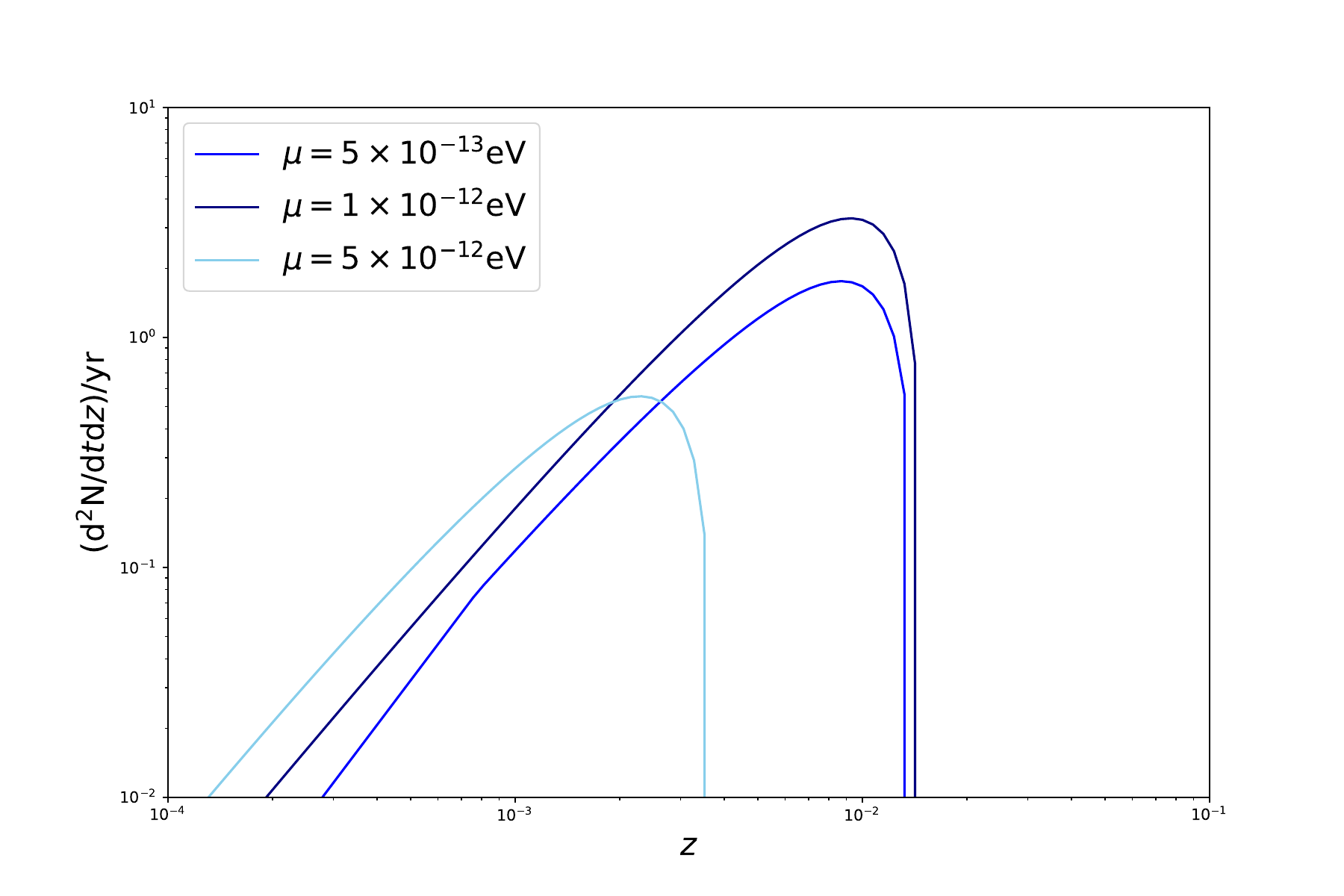}
          \caption{Event rate distribution of detectable cloud level transition GWs in redshift. We consider a BBO-like GW detector with sensitivity shown in Fig.~\ref{fig:SNR_BBO}. The blue, navy, and lighblue lines show the distributions assuming scalar fields of mass $5 \times 10^{-13} {\rm eV}$, $1 \times 10^{-12} {\rm eV}$, and $5 \times 10^{-12} {\rm eV}$, and give a total event rate of $0.015 {\rm yr}^{-1}$, $0.03 {\rm yr}^{-1}$ and $0.001 {\rm yr}^{-1}$ respectively. }
     \label{fig:rate}
 \end{figure}

Fig.~\ref{fig:rate} shows the event rate distribution of detectable cloud level transition GWs in redshift, assuming a BBO-like detector. We consider scalar fields of different masses, and find the distribution is maximized around $\mu = 10^{-12} {\rm eV}$ with a total rate of $0.03$ events per year. For fields of larger masses, the event rate decreases due to the low SNR, while for fields of smaller mass, the event rate decreases due to the lack of heavy stellar mass black holes.

\subsection{Stochastic gravitational wave background}

\begin{figure}
    \centering
    \includegraphics[width=0.9\linewidth]{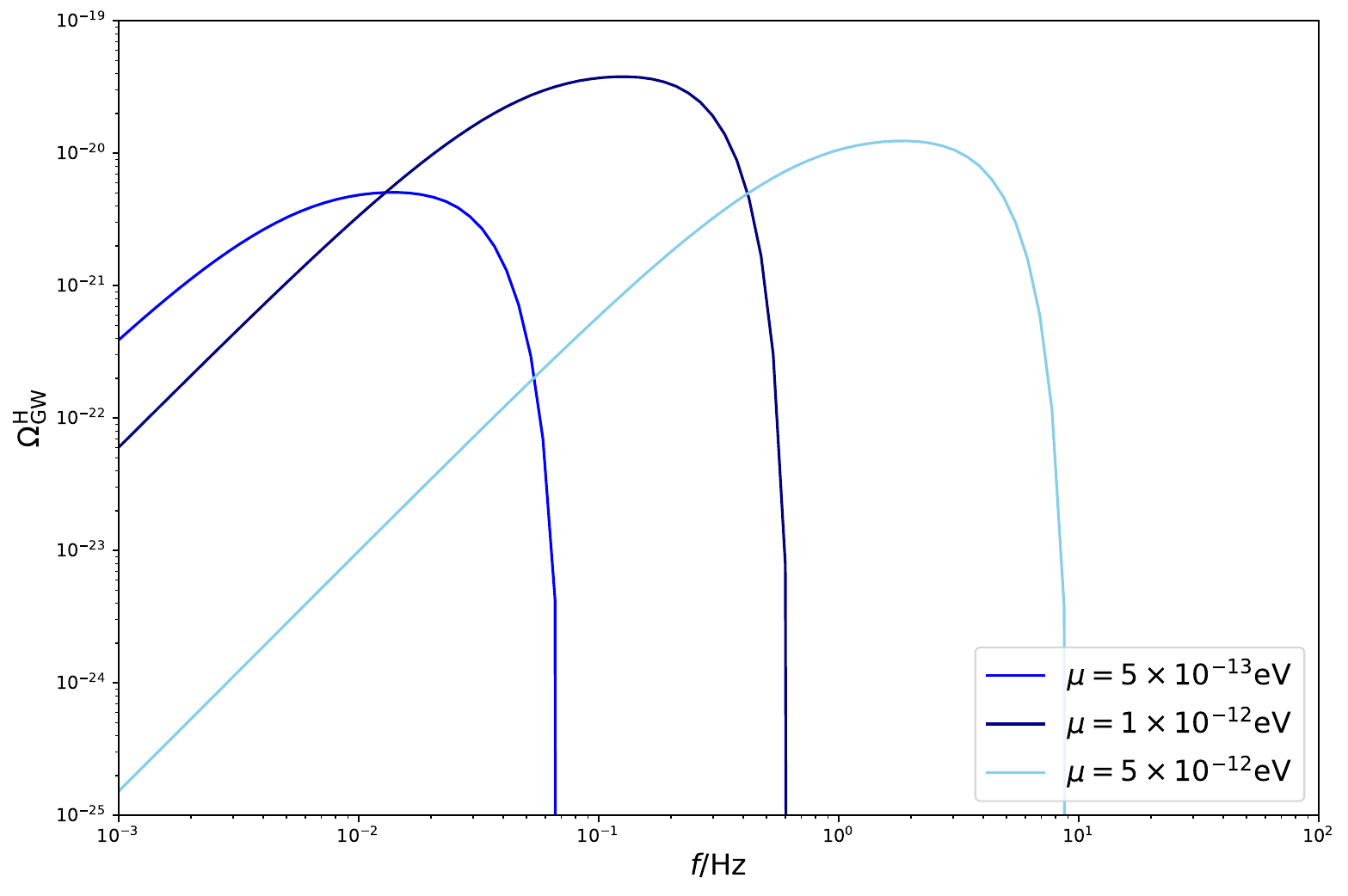}
         \caption{Energy density of the cloud level transition SGWB. We consider the hyperfine level transition between $\ket{21\pm1}$, and show the energy density of the cloud level transition SGWB of scalar fields with different scalar mass $\mu$. }
    \label{fig:sgwb}
\end{figure}

The cloud level transition GWs may also form stochastic gravitational wave background (SGWB). Different from GWs of an individual event which are typically detected with matched filtering, the SGWB is the incoherent superposition of many sources with random phases, and is typically detected by correlating multiple detectors over long observation times. The detectability of SGWB can be estimated with its energy density. Given the black hole binary population, the fractional energy density of SGWB can be calculated as
\be
    \Omega_{\rm GW}(f_o) = \frac{8 \pi G}{3 c^2 H_0^2} f_o \int d\theta p(\theta) \int dz \frac{R_{\text{mrg}}(M, t(z))}{(1+z)E_V(z)} \frac{dE(\theta)}{df} \,,
\ee
where $f_o = f / (1+z)$ is observed frequency, $E_V(z) = \sqrt{\Omega_m (1+z)^3 + \Omega_\Lambda}$ with $\Omega_m$ and $\Omega_\Lambda$ being the fractional energy density of dark matter and dark energy respectively, $\theta$ are source parameters, and $p(\theta)$ are the distributions thereof. $\frac{dE}{df}$ is the energy spectrum for a single event, and is generally given by 
\be
  \frac{dE}{df} = \frac{\pi c^3}{2 G} f^2 r^2 \int d\hat{\Omega} \left| \tilde{h} (f, \theta) \right|^2 \, ,
\ee
where the integral is over the solid angle $\hat{\Omega}$.
In our case, $\frac{dE}{df}$ only depends on the black hole mass $M$ given a certain $\mu$, and the SGWB energy density is
\be
    \Omega_{\rm GW}^{\rm H}(f_o) = \frac{8 \pi G}{3 c^2 H_0^2} f_o \int {\rm d} M \int dz \frac{R_{\text{mrg}}(M, t(z))}{(1+z)E_V(z)} \frac{dE(M)^{\rm H}_{211}}{df} \, .
\ee
In Fig.~\ref{fig:sgwb}, we show the SGWB energy density from the hyperfine level transition between $\ket{21\pm1}$, assuming scalar fields of different masses. We find that the energy density of such SGWB is typically very lower. Nevertheless, DECIGO could in principle detect SGWB with $\Omega_{GW} > 10^{-20}$ if it can reach its original design sensitivity~\cite{Seto:2001qf}, in which case the SGWB from the cloud level transition can be detected.

\section{Conclusion and Discussion}
\label{sec:con}

The detection of GWs opens up a new window to the dark sector of our universe, and have already demonstrated its potential in probing fundamental physics, e.g., see Refs.~\cite{Tsukada:2018mbp, Isi:2018pzk, Palomba:2019vxe,Tsukada:2020lgt,Ng:2020ruv,LIGOScientific:2021rnv, Yuan:2022bem,Sagunski:2017nzb,Huang:2018pbu,Sennett:2019bpc,Zhang:2021mks, Barsanti:2022vvl,Liu:2023spp}. In this work, we investigate the possibility of probing ultralight bosons with the GW signals produced in their cloud level transitions. By explicitly calculating the cloud level transition GWs, we find that, for clouds that are initially in $\ket{211}$, the hyperfine level transition can lead to quadrupole GW radiation, while fine and Bohr level transitions do not lead to significant GW radiation. Focusing on the hyperfine level transition between $\ket{21\pm1}$, we find that the resulting GWs is around $10^{-3}$ Hz for binaries with intermediate mass black holes, $0.1$Hz for stellar mass black hole binaries, and kilo-Hz for binaries of primordial black holes. We further investigate the detectability of such GWs with GW detectors such as BBO and LISA, and find that BBO could detect such GW signals at rate of 0.03 events per year if there is a scalar field with mass around $10^{-12}$ eV. We also compute the energy density of cloud level transition SGWB, which however is very low, and cannot be detected by DECIGO, assuming a 3-year observation. Our study forecasts the detectability of the cloud level transition GWs, and demonstrates such signals, if detected by a BBO-like detector, could probe ultralight bosons with mass around $10^{-12}$eV. While ultralight fields can also be probed by the quasi-monochromatic GWs emitted by clouds around isolated black holes, observing the cloud level transition GWs can help breaking the degeneracy, distinguishing the clouds from other GW sources.

While in this work we focus on the cloud level transition GWs, it is known that superradiant clouds in binaries also produce other observational signatures. In particular, the presence of clouds in binaries can alter the orbital dynamics, e.g., leading to floating and sinking orbits. These effects could also be detected by observing GWs produced by the binaries, providing another smoking gun for the existence of the clouds and hence the corresponding ultralight fields. It is also demonstrated in Ref.~\cite{Chen:2024ery}, these effects can be probed with multi-band GW observations, even without the direct detections of GWs during the resonant transitions. The prospects of probing the superradiant clouds with multi-band GW observations and with combination of all their signatures could be interesting, and are left for future studies.

~\\

\section*{ACKNOWLEDGMENTS}

J. Z. is supported by the scientific research starting grants from the University of Chinese Academy of Sciences (Grant No.~118900M061), the Fundamental Research Funds for the Central Universities (Grants No.~E2EG6602X2 and No.~E2ET0209X2), and the National Natural Science Foundation of China (NSFC) under Grant No.~12147103. 

{\bf Note added}: While we were finalizing the manuscript, a preprint~\cite{Kyriazis:2025fis} discussing similar topics appeared on arXiv.

\appendix

\bibliography{references}

\end{document}